\documentclass{PoS}

\title{Long-term multi-frequency monitoring of microquasars}

\ShortTitle{Long-term multi-frequency monitoring of microquasars}

\author{\speaker{Sergei A. Trushkin}\\
	Special Astrophysical Observatory RAS, Russia\\
	E-mail: \email{satr@sao.ru}}
\author{Nikolaj A. Nizhelskij \\
	Special Astrophysical Observatory RAS, Russia\\
	E-mail: \email{nizh@sao.ru}}
\author{Nikolaj N. Bursov \\
	Special Astrophysical Observatory RAS, Russia\\
	E-mail: \email{nnb@sao.ru}}

\abstract{
We discuss the results of the radio studies of the X-ray binaries
with relativistic jets.  We carried out a multi-frequency (1-30
GHz) daily monitoring of the radio flux variability of the microquasars
SS433, GRS1915+105, and Cyg X-3 with the RATAN-600 radio telescope
during the recent sets from 1 November 2006  to 31 August 2008.
From December 2005 radio emission of Cyg X-3 after four years relatively quiescent
levels (100-200 mJy) dropped down to $\sim$20 mJy, and
then we detected a lot of bright radio flaring events (1-20 Jy) followed
the very variable (from 0 to 0.5 crabs) 15-50 keV X-ray emission, which was monitored
in the Swift/BAT ASM program. Again from December 2007 to March 2008
we have daily measured  almost quiescent fluxes from Cyg X-3 but in April 2008
a bright radio flare with clear synchrotron self-absorption was detected.
We detected several bright short-term flares from GRS~1915+105
which could be associated with active soft X-ray events.
In intense measurements of SS433 fluxes (often mutually with X-ray and optical
observations) we detected massive ejections during powerful flares.
We discuss the various spectral and temporal characteristics of the
light curves from the microquasars. Monitoring of
the flaring radio emission is a good tracer of jet activity X-ray binaries.
}

\FullConference{VII Microquasar Workshop: Microquasars and Beyond\\
		September 1 - 5, 2008\\
		Foca, Izmir, Turkey}

\begin{document}

\section{Introduction}

The variable synchrotron radio emission is a good tracer of active
processes in relativistic jets in the Galactic X-ray binaries (XRB) --
microquasars.  We continued  the multi-frequency monitoring of SS433, Cyg~X-3
and GRS 1915+105 in order to follow the jets activity and compare the
radio light curves with X-ray light curves receiving in the ASM programs
of cosmic satellites RXTE \cite{Lev96} and Swift \cite{Bar05}.
Here we accepted flux unites:
1 crab = 75 counts/s for ASM XTE data, and 1 crab = 0.22 counts/s for
ASM Swift data.

\begin{figure}
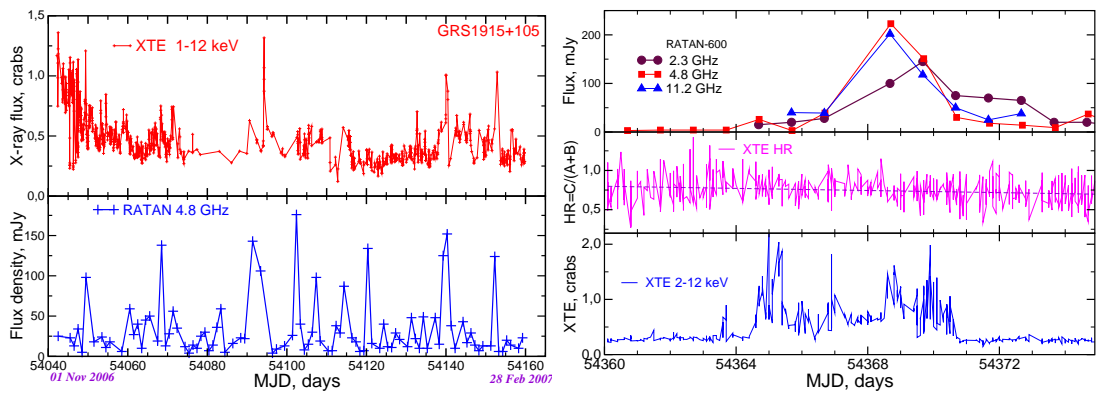

\centering
\includegraphics[width=0.48\linewidth]{1915_f1.eps}
\includegraphics[width=0.47\linewidth]{1915_f2b.eps}
\caption{{\it Left:} XTE X-ray and 4.8 GHz light curves of GRS1915+105
from 1 November 2006 to 28 February 2007. {\it Right:}
radio light curves and at XTE data:
the total flux $A+B+C$ and hardness intensity $C/(A+B)$ in September 2007.}
\label{fig:1}
\end{figure}

\section{Observations}

The all sets of the monitoring program  were carried out with North sector
RATAN-600 radio telescope on radio continuum radiometers in range 1 - 30
GHz. Usually we have daily observed three microquasars at 3-4 frequencies
(2.3, 4.8, 7.7 and 11.2 GHz), but during powerful flares we can use also 1.0
and 30 GHz radiometers with relatively worse sensitivities. The details of
the observations are given by Trushkin et al.\cite{Tru06}. In Table 1 we give
the duration of the observational sets of the microquasars during from
1 November 2006 to 1 September 2008. The last column is the total number of
daily measurements of the flux densities in the given set. Thus we have
measured more than 1000 daily spectra of microquasars during two years.
Of course we have observed 3-5 secondary calibrators daily for accurate
calibration of the measured flux densities at different frequencies.

\begin{table}
\begin{center}
\caption{Observational sets with RATAN-600 from  1 November 2006 to 1 September 2008}\vspace{1em}
\begin{tabular}{lllr}
\hline
 dd.mm.yy -- dd.mm.yy & Source      & $N_{obs}$ &\\
\hline
 03.11.06 -- 28.02.07 & V4641 Sgr   &  92 &\\
 03.11.06 -- 28.02.07 & SS433       &  99 &\\
 03.11.06 -- 28.02.07 & GRS1915+105 &  94 &\\
 03.11.06 -- 28.02.07 & Cyg X-1     &  69 &\\
 03.11.06 -- 28.02.07 & Cyg X-3     & 103 &\\
\hline
 01.06.07 -- 07.06.07 & Cyg X-3     &   5  &\\
 12.06.07 -- 04.07.07 & SS433       &  11  &\\
 21.06.07 -- 04.07.07 & GRS1915+105 &   9  &\\
\hline
 13.09.07 -- 05.10.07 & GRS1915+105 &  22  &\\
 29.09.07 -- 05.10.07 & SS433       &   7  &\\
\hline
 30.10.07 -- 18.11.07 & SS433       &  18  &\\
 30.10.07 -- 18.11.07 & GRS1915+105 &  18  &\\
\hline
 01.12.07 -- 03.03.08 & LS 5039     &  92 &\\
 01.12.07 -- 03.03.08 & SS433       &  78 &\\
 01.12.07 -- 03.03.08 & GRS1915+105 &  78 &\\
 01.12.07 -- 03.03.08 & Cyg X-1     &  76 &\\
 01.12.07 -- 03.03.08 & Cyg X-3     &  76 &\\
\hline
 08.03.08 -- 22.05.08 & SS433       &  54 &\\
 07.04.08 -- 22.05.08 & GRS1915+105 &  44 &\\
 13.04.08 -- 27.05.08 & Cyg X-3     &  10 &\\
\hline
 17.06.08 -- 31.08.08 & SS433       &  68 &\\
 17.06.08 -- 31.08.08 & GRS1915+105 &  61 &\\
\hline
\end{tabular}
\label{tab:1}
\end{center}
\end{table}

\section{Discussion}

\subsection{GRS 1915+105: X-ray -- radio correlation}

The soft/hard X-ray - radio flaring correlation of GRS1 1915+105 was discussed
a lot of times \cite{Fos96,Han99,Fen02,Vad03,Yad06}.
In Fig.\ref{fig:1}(left) the 4.8GHz and X-ray light curves are
shown during the set in 2006-2007. The frequent bright radio flares have
the associated bright events from X-ray light curves received in ASM RXTE
at 2-12 keV.

Fig.\ref{fig:1}(right) shows radio light curves of the bright flare and at
XTE data: the total flux $S=A+B+C$ (1.5-3, 3-5, 5-12keV) bands of ASM XTE and the X-ray hardness
ratio $HR=C/(A+B)$ in the set of September 2007.  Clearly that radio flare
associated with X-ray event and we did not detected any changes in HR as
were discussed by Namiki et al. \cite{Nam06} and Trushkin \cite{Tru07a,Tru08a}. The radio spectrum
was optically thick at frequencies lower 4.8 GHz during the flare.

\begin{figure}
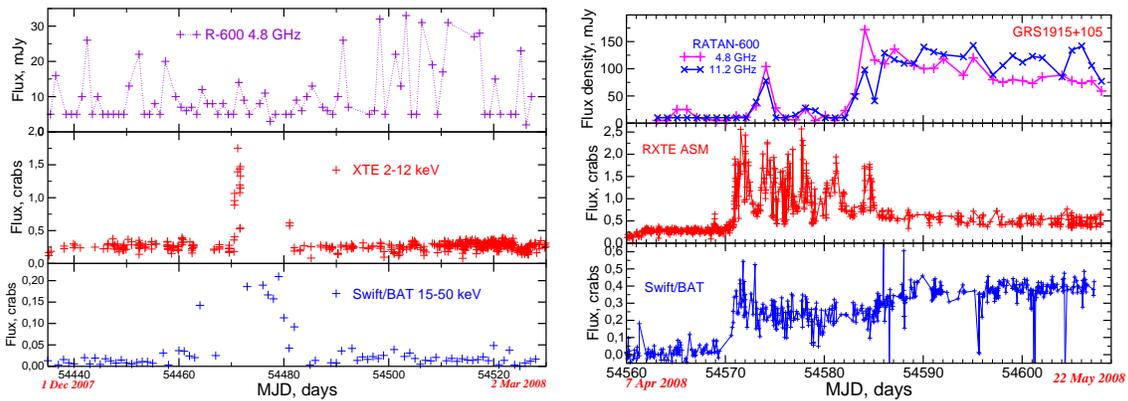

\centering
\includegraphics[width=0.48\linewidth]{1915_f3.eps}\hspace{0.3cm}
\includegraphics[width=0.48\linewidth]{1915_f4.eps}
\caption{{\it Left:} Light curves of GRS1915+105 at 4.8 GHz/11.2GHz, at 2-12 keV
and 15-50 keV from 1 December 2007 to 2 March 2008. {\it Right:}
the same from 7 April to 22 May 2008.}
\label{fig:2}
\end{figure}

In Fig.\ref{fig:2}(left) the light curves of GRS1915+105 at 4.8 GHz, at
2-12 keV and  at 15-50 keV bands from 1 December 2007 to 2 March 2008
are shown. GRS 1915+105 was in quiet state, and unusual changes of X-ray
fluxes near MJD54470 are probably connected with close solar elongation
of the source. Thus the microquasar has fluxes lower than 30 mJy and
was relatively stable in all bands.

But already in April 2008 GRS1915+105 passed in active state with high
levels of sort and hard X-ray emission. After that we detected two
bright radio optically thin flares and then the source stayed on fluxes
about 100 mJy with flat or even inverted radio spectra, as shown in
Fig.\ref{fig:2}(right).

\begin{figure}
\centering
\includegraphics[width=0.8\linewidth]{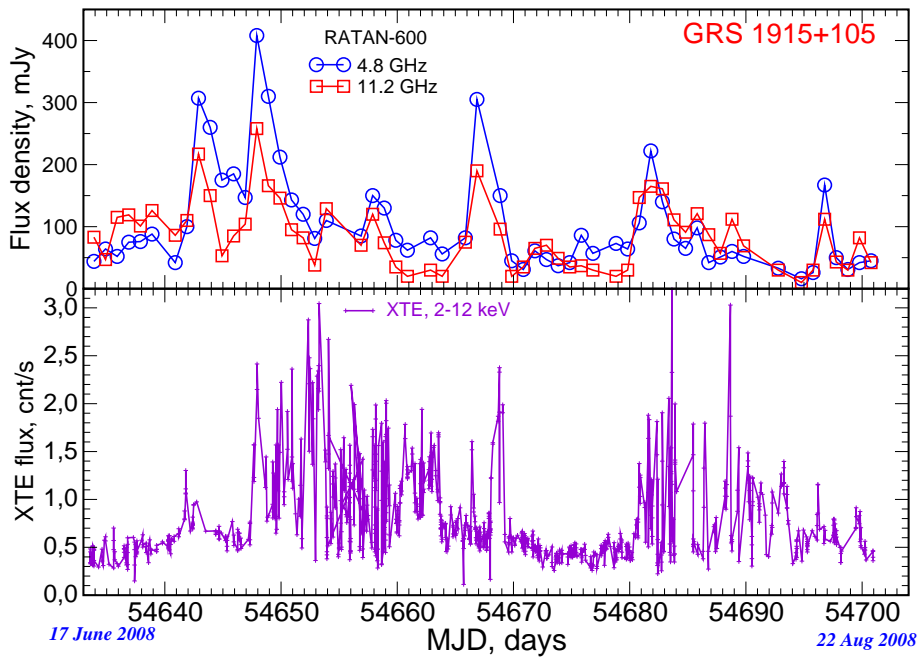}
\caption{Light curves of GRS1915+105 at radio frequencies, at 2-12 keV
from 17 June to 22 Aug 2008.}
\label{fig:3}
\end{figure}

Al last during summer months 2008 GRS1915+105 was very active and four flare
with peak fluxes higher 200 mJy were detected (Fig.\ref{fig:3}).
Again these bright flares have the associated bright soft X-ray events
at 2-12 keV.

\subsection{Cyg X-3: 2006 -- new long-term active period }

Recently Szostek et al. \cite{Szo08} discussed the long-term correlations of
the X-ray and radio states, using BATSE, XTE and GBI (\cite{Wal94,Wat94} data.
In \cite{Szo08} was presented the detailed classification of the mutual states.

\begin{figure}
\centering
\includegraphics[width=0.7\linewidth]{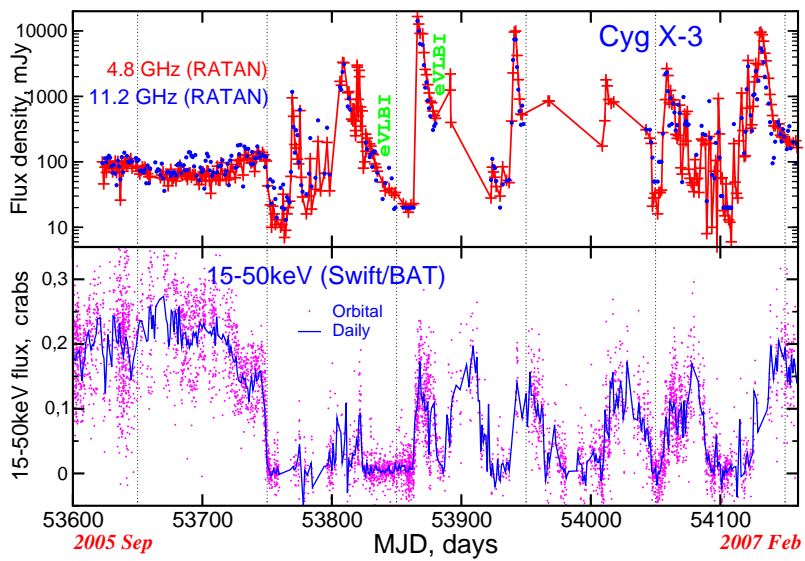}
\caption{The RATAN 4.8/11.2 GHz and Swift/BAT ASM light curves of Cyg X-3
from  September 2005 to February 2007.}
\label{fig:4}
\end{figure}

\begin{figure}
\centering
\includegraphics[width=0.7\linewidth]{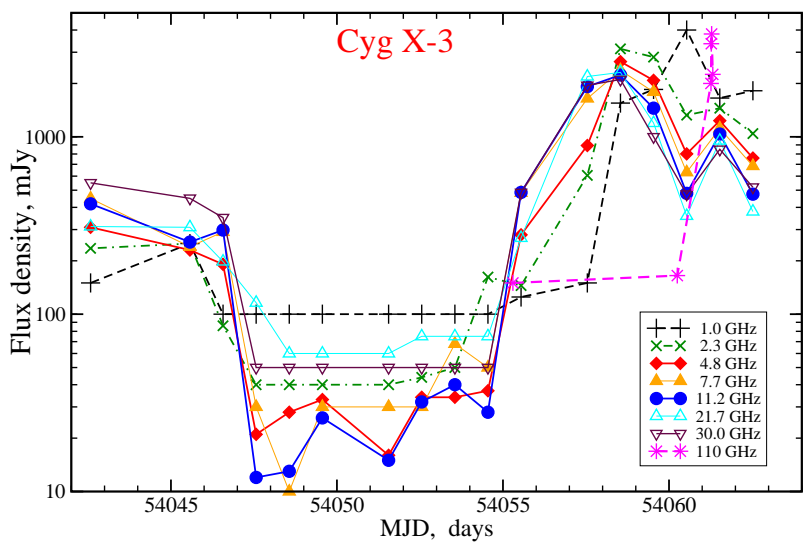}
\caption{Radio light curves of Cyg X-3 before and during the bright
flare in November 2006.}
\label{fig:5}
\end{figure}

\begin{figure}
\centering
\includegraphics[width=0.7\linewidth]{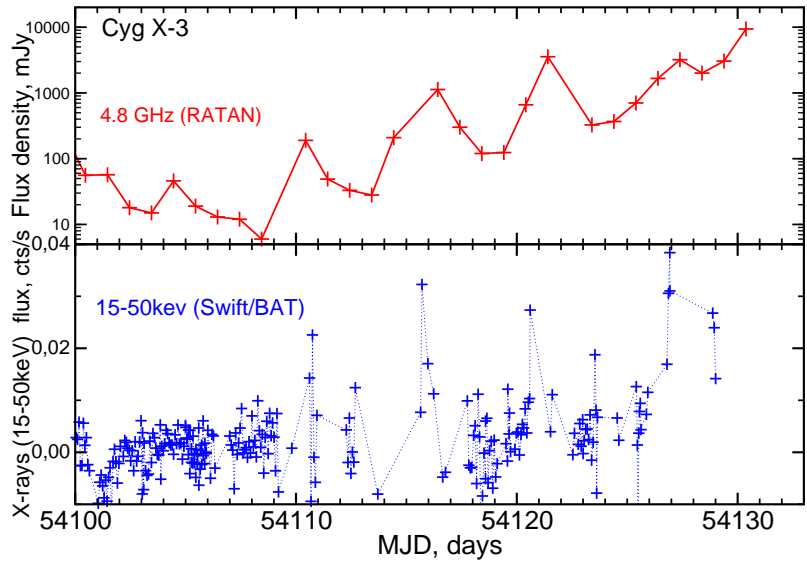}
\caption{The radio and X-ray light curves of the Cyg X-3 during flare
in January 2007.}
\label{fig:6}
\end{figure}

\begin{figure}
\centering
\includegraphics[width=0.7\linewidth]{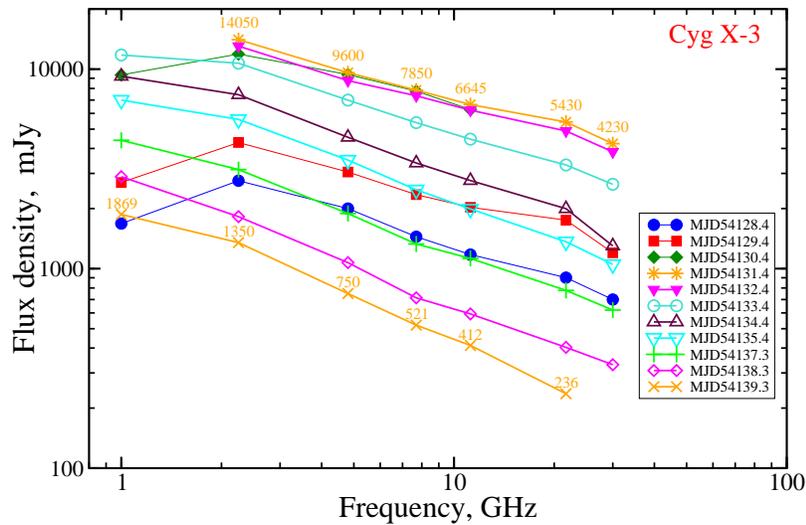}
\caption{The daily spectra of Cyg X-3 during the flare in February 2007.}
\label{fig:7}
\end{figure}

\begin{figure}
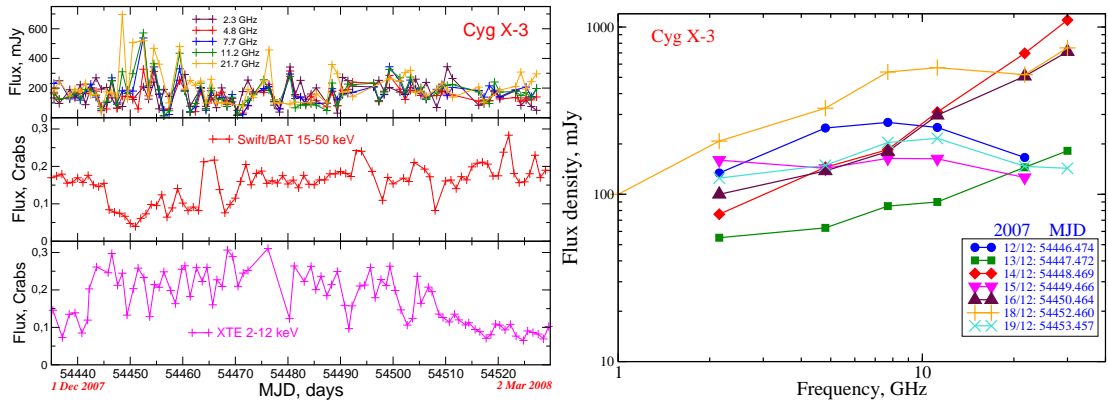

\centering
\includegraphics[width=0.48\linewidth]{c3_f5.eps}
\includegraphics[width=0.48\linewidth]{c3_f5s.eps}
\caption{ {\it Left:} The radio and X-ray light curves of Cyg X-3 from 1 December 2007 to 2 March 2008.
{\it Right:} the daily spectra of Cyg X-3 during flaring events in December 2007.}
\label{fig:8}
\end{figure}

\begin{figure}
\centering
\includegraphics[width=0.47\linewidth]{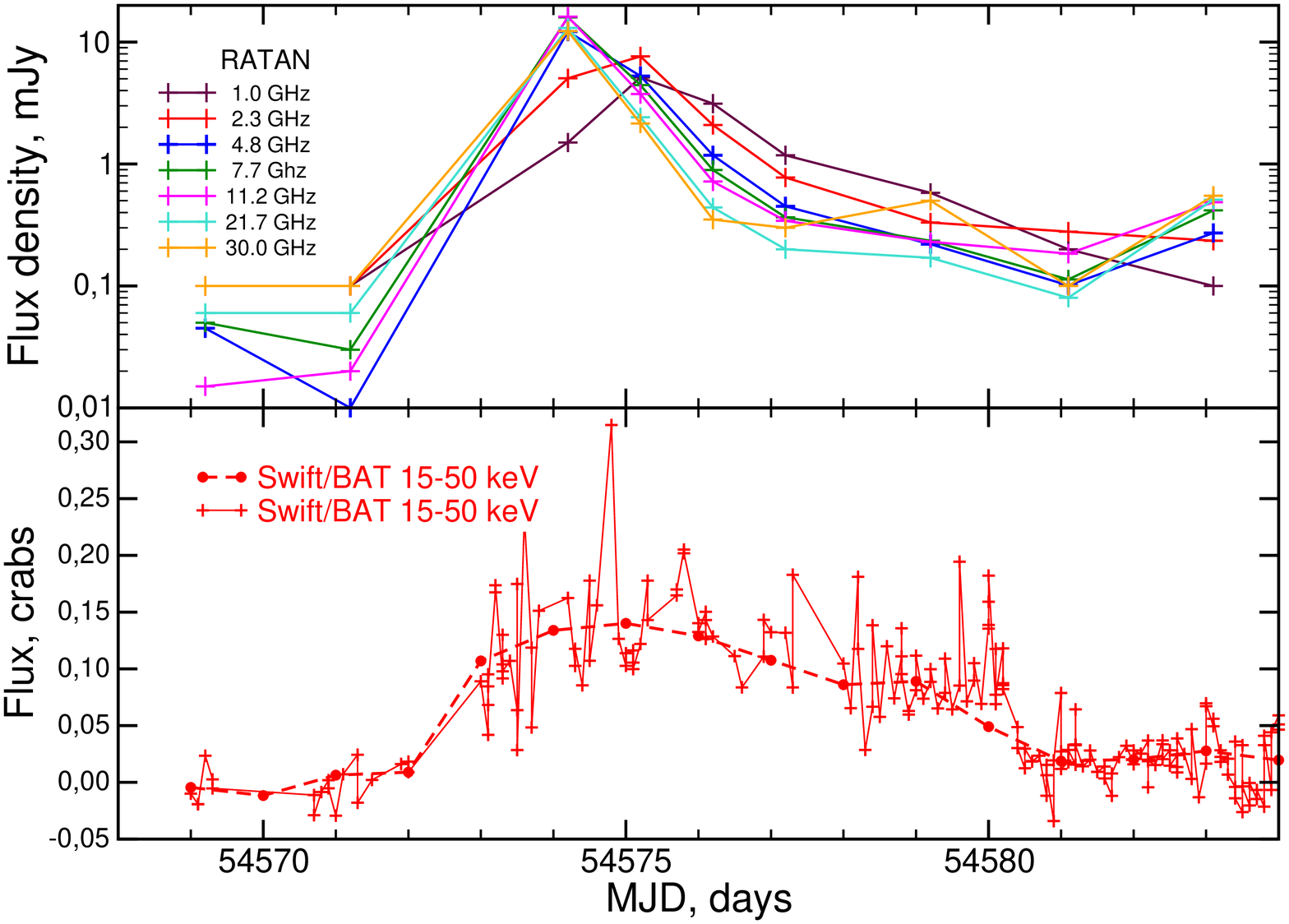}
\hspace{0.2cm}
\includegraphics[width=0.49\linewidth]{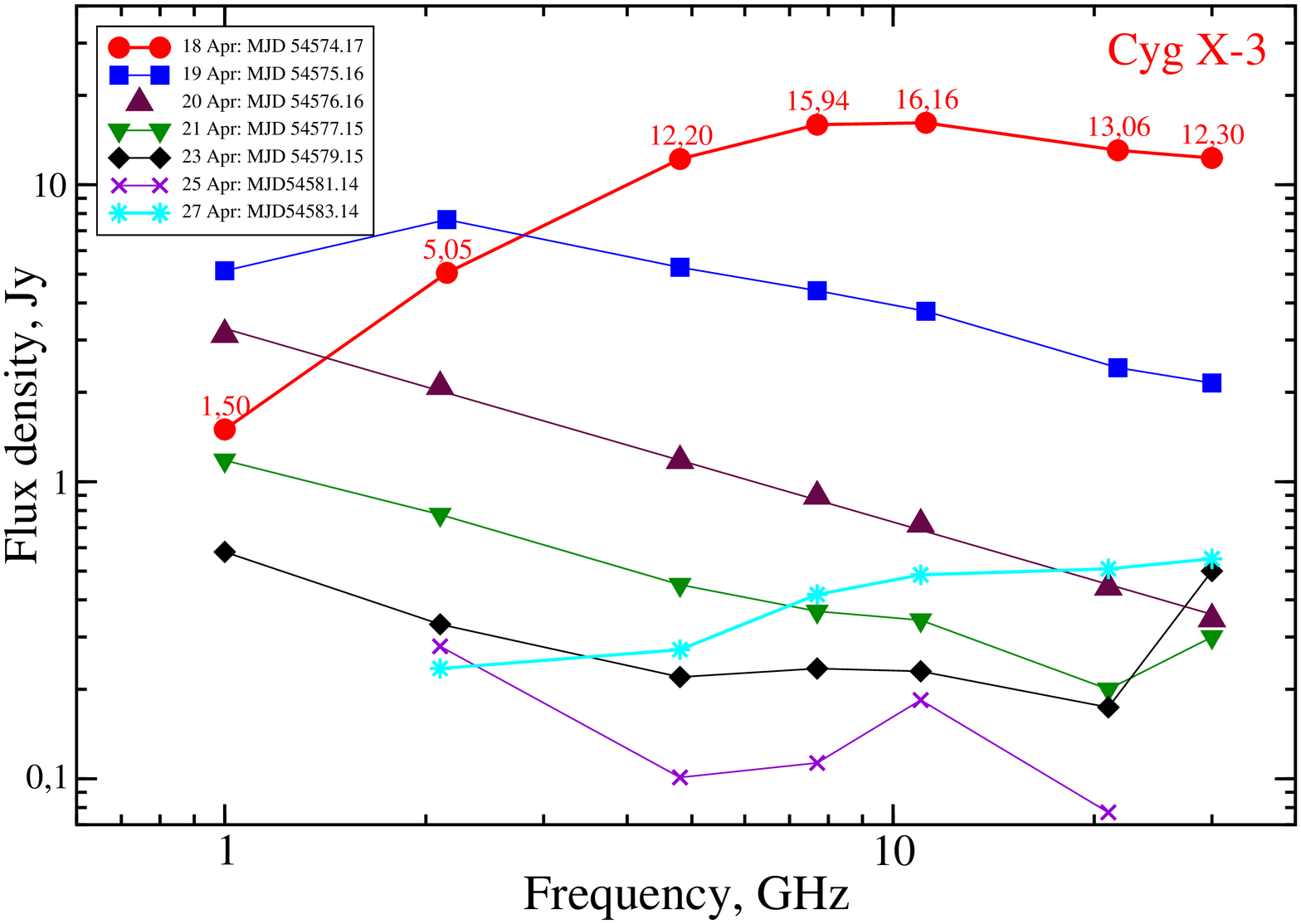}
\caption{{\it Left:} The RATAN and Swift light curves of Cyg X-3 in April 2008. {\it Right:}
the daily spectra of Cyg X-3 during the flare in April 2008.}
\label{fig:9}
\end{figure}

Trushkin et al.\cite{Tru06,Tru08a} and Tsuboi et al.\cite{Tsu06,Tsu08} discussed the
beginning of the active period of Cyg X-3 in  December 2005 -- March 2006.
The activity continued to 2008 (Fig.\ref{fig:4}) and a lot of strong
flares were  detected in March, May and Jule 2006, with peak fluxes
3-5, 12-18 and 8-12 Jy respectively. In the May 2006 flare fluxes have
grown up by a factor $\sim$1000 during a day. Just 20 April and 18 May were
received the first eVLBI maps of Cyg X-3 \cite{Tud07}.

In November 2006 we found new evidence of the activity of Cyg X-3 with
a flare typically with peak exceeding 3-5 Jy following a quenched state
($<20$mJy)(Waltman et al. \cite{Wal94}).  In Fig.\ref{fig:5} the
milti-frequency RATAN light curves are given with some observations
with Nobeyama Radio observatory interferometer (NMA).  We detected very
variable mm-band (110 GHz) emission after maxima at lower frequencies,
which was evolved from 150 mJy to 3 Jy and then 2 Jy during 1-2 hours. The
RATAN spectrum on Nov 22.53UT (MJD54061.5) is well fitted by a power-law
with spectral index -0.19 from 1 to 30 GHz. The total spectrum from 1
GHz to 100 GHz on Nov 22 can not be expressed with an optically thin
power-low model but shows a high-frequency component with a peak above
100 GHz and/or with a short lifetime, less than 6-8 hours.  The mm-band
flare have probably weaker counterparts at the lower RATAN-600 frequencies
(Tosaki el al.\cite{Tos06}). Before we have detected the fluxes changed from $\sim1$
Jy to 2 Jy during 3 hours and then decreased to 100-400 mJy during 15
hours with at 2.3 and 8.5 GHz with RT32 (IAA) on 5 June 2006 (MJD53891)
\cite{Tru06b}.

In Fig.\ref{fig:6} the 4.8 GHz and Swift 15-50 keV light curves
of the beginning of January 2007 flare are shown. We can clearly see that
all  four radio small flares have X-ray counterparts -- short-time spikes,
just before ($\sim1$day) radio local maxima.

After this increase stage in January 2007 we plotted the daily spectra
of Cyg X-3 in February 2007 (Fig.\ref{fig:7}).  A a rule only during
first drop-up stage their spectra have turn-overs at frequencies,
lower than 2 GHz.  The spectral and temporal evolution  of such flares
could be successfully fitted with the modified finite segments model
by Hjellming et al.\cite{Hj00} and developed
by Marti et al. \cite{Mar92}. Trushkin et al. \cite{Tru06,Tru08a} have fitted spectra
of the July 2007 flare. In the first stage of such a flare we should
involve the intense internal shocks running through the jet as proposed
Watanabe et al. \cite{Wat94}. The former flare in September 2001 was successfully
fitted  by Lindfords et al. \cite{Lin07}.

In Fig.\ref{fig:8}(left) the radio (RATAN) and X-ray light curves
(XTE, Swift) of Cyg X-3 from 1 December 2007 to 2 March 2008 are shown.
Only in the beginning of the set Cyg X-3 was in active state with relatively
small hard X-ray emission (MJD54450). Then spectra of Cyg X-3 were
inverted with positive spectral indices, as shown in Fig.\ref{fig:8}(right).

In April 2008 we detected the bright flare from Cyg X-3 again after a
quenched state.  In Fig.\ref{fig:9}(left) the radio (RATAN) and X-ray
light curve (Swift/BAT) of the flare are shown.  In Fig.\ref{fig:9}(right)
the daily radio spectra are plotted.  The 18 April spectrum give clear
evidence of the optically thick regime at lower frequencies. With GRMT
points at 614 and 244 MHz \cite{Pal06,Pal08} the quasi-simultaneous spectra could be well
fitted by or a synchrotron self-absorption model, when spectral index is equal to -2.5
or a external absorbed screen model. The turn-over frequency is shifted towards lower frequency
day by day.

\begin{figure}
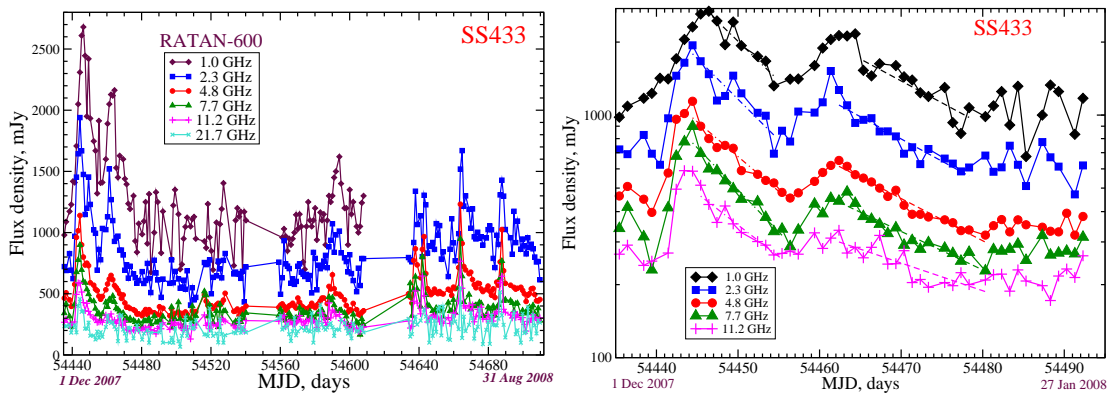

\centering
\includegraphics[width=0.48\linewidth]{ss433_f1.eps}
\includegraphics[width=0.48\linewidth]{ss433_f2.eps}
\caption{ {\it Left:} The RATAN and Swift light curves of SS433 from 1 December 2007 to 31 August 2008.
{\it Right:} radio light curves  from 1 December 2007 to 22 January 2008 in log scale (right).}
\label{fig:10}
\end{figure}

\subsection{SS433: a intense set in 2007-2008}

We continue  studies multi-frequency variability of the first microquasar
SS433 in order to detect the massive ejections in the flaring states \cite{Kot06,Kot07,Kot08}.
Many monitoring sets (e.g., with GBI (\cite{Fie87}),
RATAN (Trushkin et al.\cite{Tru03}) were began in 1970-80ths.
In Fig.\ref{fig:10}(left) the light curves of SS433 during
1 December 2007 - 31 August 2008 are shown.

In Fig.\ref{fig:10}(right) the light curves during the bright flares
in December 2007 - February 2008 are shown.  The both flares decayed
with exponential law ($S_\nu \propto exp(-t/t_0)$, where $t_0 = 16$days
and $t_0 = 25$days for the both flares respectively.  The delay of the
maximum flux of the bright flare in December 2007 at 1 GHz is about 2
days and 1 day at 2.3  GHz relative to the maxima at higher frequencies.
We detected the surprising coincident dates (6-7 Dec) of the bright
flares in 2006 and 2007.  Probably the 1-year periodicity of activity
exit in radiation of SS433. Recently Nandi et al. \cite{Nan05} discussed the
periodicity of flaring events in ASM RXTE X-ray data, and found possible
period about 368 days.

\section{Conclusions}

In 2006-2008 the RATAN microquasar monitoring data (1200 radio spectra)
give us abundant material for detailed comparison with the X-ray data from
the ASM or ToO programs with RXTE, CHANDRA, Suzaku and INTEGRAL.
The 1-30 GHz emission originates often
from different optically thin and thick regions. That could give us
a key for adequate modelling of the flaring radio radiation formed in
the relativistic jets interacting with varying circumstellar medium or
stellar winds.

{\it Acknowledgments}.
These studies are supported by the Russian  Foundation  Base Research (RFBR)
grant N~08-02-00504 and travel grant N~08-02-08711.

\end{document}